\documentclass[12pt,a4paper]{article}
  \usepackage{a4wide}
  \usepackage{latexsym}
  \usepackage{epsf}
  \usepackage{amssymb}
  \usepackage{amsmath, cite}
  \usepackage{slashed,epsfig}
  \usepackage{amsfonts}
  \usepackage{bbm}
  \usepackage{hyperref}
 \usepackage{verbatim}
 \usepackage{amsmath,amssymb,amsthm}

\newcommand{\Real}{\mathbb{R}}
\newcommand{\e}{\textrm{e}}
\renewcommand{\d}{\textrm{d}}

\def\be{\begin{equation}}
\def\ee{\end{equation}}
\def\bea{\begin{eqnarray}}
\def\eea{\end{eqnarray}}

\def\a{\alpha}

\def\call         {{\cal L}}

\def\calq         {{\cal Q}}

\def\calv         {{\cal V}}
\def\calw         {{\cal W}}

\begin{document}
\numberwithin{equation}{section}
\begin{flushright}
\small
IPhT-T12/019\\
%
\normalsize
\end{flushright}
\vspace{0.8 cm}

\begin{center}
{\LARGE \textbf{Fake supersymmetry versus Hamilton--Jacobi}}

\vspace{1.2 cm} {\large Mario Trigiante$^{a}$, Thomas Van Riet$^{b}$, Bert Vercnocke$^{b}$}\\

\vspace{1cm} {$^{a}$
Laboratory of Theoretical Physics,\\
Department of Applied Science and Technology, Politecnico di Torino, \\
C.so Duca degli Abruzzi, 24, I-10129 Torino, Italy}

\vspace{0.5cm} {$^{b}$ Institut de Physique Th\'eorique, CEA Saclay,\\
CNRS URA 2306,  F-91191 Gif-sur-Yvette,
France}

\vspace{.5cm}
mario.trigiante @ gmail.com, thomas.van-riet,
bert.vercnocke @ cea.fr\\

\vspace{2cm}

\textbf{Abstract}
\end{center}

\begin{quotation}
We explain when the first-order Hamilton--Jacobi equations for black
holes (and domain walls) in (gauged) supergravity, reduce to the
usual first-order equations derived from a fake superpotential. This
turns out to be equivalent to the vanishing of a newly found
constant of motion and we illustrate this with various examples. We
show that fake supersymmetry is a necessary condition for having
physically sensible extremal black hole solutions. We furthermore
observe that small black holes become scaling solutions near the
horizon. When combined with fake supersymmetry, this leads to a
precise extension of the attractor mechanism to small black holes:
The attractor solution is such that the scalars move on specific
curves, determined by the black hole charges, that are purely
geodesic, although there is a non-zero potential.
\end{quotation}
\newpage

\tableofcontents

\section{Introduction}
Supersymmetry has proven to be a very powerful tool in string
theory. In the supergravity approximation,
supersymmetry can be used as a tool to generate solutions since
supersymmetric solutions obey first-order equations that are derived
from the preservation of some fraction of supersymmetry. However,
the ultimate goal is to perform computations in string theory in
non-supersymmetric contexts. For the sake of finding supergravity
solutions, such as black holes for instance, the so-called
fake supergravity formalism borrows tricks of supersymmetry to
find non-supersymmetric solutions. Fake supergravity was first
introduced in the context of domain wall solutions in
\cite{Freedman:2003ax}. For supersymmetric domain wall solutions, the
first order supersymmetry equations are determined by the superpotential.
The idea of fake supergravity (or fake supersymmetry) is to define a fake superpotential,
not directly related to the superalgebra, whose
first-order gradient flow also satisfies the second-order equations
of motion. Apart from being a technical tool, fake supersymmetry has
a physical meaning in the sense that it guarantees the stability of
the domain wall solution. It thus becomes important to understand
when a domain wall solution is fake supersymmetric or not, a
question which was raised in \cite{Celi:2004st}.

FLRW solutions supported by time-dependent scalars subject to the
force derived from some scalar potential are very similar to
gravitational domain wall solutions. In fact there exists a one to
one map between such solutions \cite{Skenderis:2006jq}. This implies
that the notion of fake supersymmetry should carry over to
time-dependent solutions. This is called pseudo-supersymmetry since
time-dependent solutions cannot possess unitary
superalgebras.\footnote{ For earlier remarks on the first-order
formalism in cosmology we refer to \cite{Liddle:2000cg,
Bazeia:2005tj} and for applications to bent branes, see
\cite{Afonso:2006gi}.}

Domain walls and FLRW cosmologies are solutions that depend on
one coordinate. The second-order equations of motion take the form
of ODE's that can be derived from an effective Lagrangian
that defines a Hamiltonian system. This suggests a close link between
fake supergravity and the Hamilton--Jacobi formalism, since the
latter also defines first-order equations. Indeed such a link was
established in \cite{Skenderis:2006fb} (see also
\cite{Skenderis:2006rr, Townsend:2007aw, Townsend:2007nm}).
Nonetheless, it has never been precisely formulated what the
difference is between fake supersymmetry and the Hamilton--Jacobi formalism. It is
the aim of this paper to make this precise. It is important to note
that not all domain walls are derived  from fake supersymmetry. Reference
\cite{Sonner:2007cp} gave an explicit example for which one can
easily proof that there is no fake supersymmetry in the standard
sense. We will recall and simplify this solution in this paper and
present even simpler examples.

Spherically symmetric stationary black holes are another example of co-homogeneity one
solutions in supergravity that effectively define a Hamiltonian
system. It is therefore not surprising that the fake supergravity
formalism can be applied to such black holes as well. This was
initiated in \cite{Ceresole:2007wx} and many other references soon
followed that constructed fake superpotentials for non-supersymmetric extremal
black holes (see e.g.~\cite{Lopes Cardoso:2007ky,
Andrianopoli:2007gt,Ceresole:2009vp} and the review \cite{Dall'Agata:2011nh} for
more references). Similarly to domain walls, the relation with the
Hamilton--Jacobi formalism was noted in \cite{Andrianopoli:2009je}.
Interestingly the existence of first-order gradient flow equations
was then also found for non-extremal black holes in Einstein
Maxwell-theory \cite{Miller:2006ay} and extended to more general
models in \cite{Janssen:2007rc}. The most general form for the
first-order equations for non-extremal solutions was found in
\cite{Perz:2008kh}, where it was emphasized that the flow equations
do differ from those for extremal black holes in the sense that the black hole
warp factor appears in a non-trivial way, different from extremal
solutions.

The only physical interpretation of the fake superpotential for black holes
so far is as a Liapunov function \cite{Andrianopoli:2010bj}. A deeper physical
meaning of the existence of fake supersymmetry, similar to the
assurance of stability for domain walls, has not been understood.
One of the purposes of this paper is to fill this gap for extremal
black holes.


This paper is organized as follows.
We recall  the concept of fake supersymmetry in section \ref{sect:2}
and the Hamilton--Jacobi (HJ) formalism in section \ref{sect:3}. In section \ref{sect:4},
we show that the HJ equations reduce to the standard fake supersymmetry equations
if a certain, newly-found, constant of motion vanishes. We furthermore give a new
physical interpretation of fake supersymmetry in the context of extremal black holes: fake
supersymmetry is necessary for having physically acceptable
solutions. We use this in section \ref{sect:5} to uncover new general properties
of the near horizon regions of small black holes, which can be
regarded as an extension of the attractor mechanism for large
extremal black holes \cite{Ferrara:1995ih, Ferrara:1996dd}. We end with a discussion in
section \ref{sect:6}. Appendix \ref{sec:Examples} contains several examples
that clarify the statements in the bulk of the paper.

\section{Fake supersymmetry}\label{sect:2}
Fake supersymmetry is a concept that is usually formulated on the
level of effective one-dimensional actions for black hole solutions,
domain walls and, through the map between domain walls and FLRW
cosmologies \cite{Skenderis:2006fb}, also for the latter, where it
is referred to as pseudo-supersymmetry. We briefly recall these
effective actions before we recall the notion of fake supersymmetry.

\subsubsection*{Effective actions for black holes} A typical ungauged
supergravity is described by $N$ real scalar fields $\phi^i$ that
parameterize a Riemannian target space with metric $G_{ij}$ and
$M$ Abelian vector fields that couple to these scalars. Spherical
and static black hole solutions to such theories are described by
the following Ansatz
\begin{equation}
\d s^2_4 = -\e^{2U(\tau)}\d t^2 +
\e^{-2U(\tau)}\Bigl(\e^{4A(\tau)}\d\tau^2 +
\e^{2A(\tau)}\d\Omega^2_2\Bigr)\,.
\end{equation}
In the absence of a scalar potential (ungauged supergravity) the
function $A(\tau)$ is independent of the matter content:
\begin{align}
& \text{extremal:} \qquad \qquad \e^{A(\tau)} = \frac 1 \tau\,,\\
& \text{non-extremal:} \quad\quad \e^{A(\tau)} =
\frac{c}{\sinh(c\tau)}\,.
\end{align}
The constant $c$ is the so-called non-extremality parameter and
$\tau$ is a reparametrisation of the usual radial coordinate.
When $c^2<0$, we find non-physical (`over-extremal') solutions
(although the metric is still real for some coordinate range).

Due to spherical symmetry the scalars only depend on the radial
direction and the vector fields can easily be integrated out
in terms of the magnetic and electric charges. The equations of
motion are captured by the effective action
\begin{equation}\label{BHeffectiveaction}
S=\int \d\tau\Bigl(4\dot{U}^2 + G_{ij}\dot{\phi}^i\dot{\phi}^j  -
\e^{2U}V_{BH}(\phi)\Bigr)\,,
\end{equation}
where a dot represents a derivative with respect to $\tau$. The
black hole potential $V_{BH}(\phi)$ is the term generated by
integrating out field strengths and  is strictly negative.
The solutions are subject to the following energy
constraint
\begin{equation}\label{energy3}
G_{ij}\dot{\phi}^i\dot{\phi}^j + 4\dot{U}^2 + \e^{2U}V_{BH}=4c^2\,.
\end{equation}

\subsubsection*{Effective actions for domain walls}
Domain wall solutions are usually supported by scalar fields that
are subject to a non-trivial scalar potential $V_{DW}(\phi)$. The
standard Ansatz for flat domain wall solutions is given by
\begin{equation}
\d s_2^2 = \e^{2\sqrt3 U(z)}\d z^2 +
e^{\tfrac{2}{\sqrt3}U(z)}\Bigl(\d x^2 + \d y^2 -\d t^2 \Bigr)\,.
\end{equation}
Consistent with the symmetries one then typically assumes that the
scalars depend on $z$ only. The effective action then becomes
one-dimensional
\begin{equation}\label{DWeffectiveaction}
S=\int \d z\Bigl( 4\dot{U}^2 -G_{ij}\dot{\phi}^i\dot{\phi}^j -
\e^{2\sqrt3 U}V_{DW}(\phi)\Bigr)\,,
\end{equation}
and is supplemented with a zero energy condition
\begin{equation}\label{}
4\dot{U}^2 -G_{ij}\dot{\phi}^i\dot{\phi}^j + \e^{2\sqrt3
U}V_{DW}(\phi)= 0\,.
\end{equation}
Up to signs and factors of $\e^U$ there is no difference between
extremal black holes and domain walls from the point of view of the
effective action.

There is a simple one to one map between domain walls and FLRW cosmologies.
For Minkowski-sliced domain walls and $k=0$ FLRW
cosmologies, given by the metric,\footnote{We have chosen a similar $z$
parametrization as for domain walls rather than
the usual cosmological time.}
\begin{equation}
\d s_2^2 = -\e^{2\sqrt3 U(t)}\d t^2 +
e^{\tfrac{2}{\sqrt3}U(t)}\Bigl(\d x^2 + \d y^2 +\d z^2 \Bigr)\,,
\end{equation}
the map proceeds by flipping the sign of the scalar potential $V$
and replacing $z \rightarrow t$ in $U(z), \phi^i(z)$.

\subsubsection*{The fake superpotential}
Supersymmetric domain walls and black holes are a special subset of
solutions that fulfill certain first-order differential equations
that follow from the Killing spinor equations. These equations take
the form of flow equations, derived from the superpotential function
$W$:
\begin{equation}\label{floweqs}
\dot{\phi}^i=\epsilon \e^{aU} G^{ij}\partial_jW(\phi)\,,\qquad
4\dot{U}=a\e^{aU}W(\phi)\,.
\end{equation}
where
\begin{align}
&\text{Domain walls}: \qquad\epsilon=-1\qquad a=\sqrt3\nonumber\,,\\
&\text{Black holes}\quad :\qquad \epsilon=+1\qquad a=1\nonumber\,.
\end{align}
As is well known, supersymmetric black holes are necessarily
extremal. Hence the zero energy condition for both domain walls and
black holes implies a relation for the superpotential function $W$
\begin{equation}\label{Wequation}
 \epsilon\, G^{ij}\partial_i W\partial_j W + \frac{a^2}{4}W^2 =
 -V(\phi)\,.
\end{equation}

For supersymmetric solutions, $W$ is the superpotential derived from the
Killing spinor equations. The essence of fake supersymmetry is that
\emph{any} function $W$ that obeys the above relation (\ref{Wequation}) defines a first-order
flow, through (\ref{floweqs}), that can easily be demonstrated to
solve the full second-order equations of motion. Hence solutions
that can be found from a flow governed by a fake superpotential, in a
certain sense mimic  supersymmetric solutions.

\section{Hamilton's principal function}\label{sect:3}
To introduce the Hamilton--Jacobi formalism in this context, we
first write the above effective actions (\ref{BHeffectiveaction},
\ref{DWeffectiveaction}) in a more formal Hamiltonian system
notation. For that we define the configuration space variables
$q^a=(U, \phi^i )$ and the corresponding metric
\begin{equation}\label{metric}
G_{ab}=\begin{pmatrix} 4 & 0 \\ 0& \epsilon G_{ij}
\end{pmatrix}\,.
\end{equation}
We furthermore define
\begin{equation}\label{potential}
\calv(q)=\tfrac{1}{2}\e^{2aU}V(\phi)\,,
\end{equation}
then the effective action is compactly written as
\begin{equation}\label{effectiveaction}
S=\int \d\tau \Bigl(\tfrac12 G_{ab}\dot{q}^a\dot{q}^b -
\calv(q)\Bigr)\,.
\end{equation}
This action is scaled with a factor $\tfrac{1}{2}$ with respect to
the effective actions (\ref{BHeffectiveaction},
\ref{DWeffectiveaction}). The Hamiltonian is given by
\begin{equation}
\mathcal{H}=\tfrac12 G_{ab}\dot{q}^a\dot{q}^b + \calv(q)\,,
\end{equation}
which is zero for extremal black holes and domain walls and equal to
$2c^2$ for non-extremal black holes. Now we follow the steps of \cite{Andrianopoli:2009je}. The canonical
momenta $p^a$ are defined via
\begin{equation}
p_a=G_{ab}\dot{q}^b\,.
\end{equation}
Let us assume that there exists a local Hamilton--Jacobi (HJ)
formulation. We will later comment on this assumption. HJ implies the
existence of new variables $P,Q$ that obey the following equations
in terms of the \emph{principal Hamiltonian function} (or Hamilton's
principal function) $S(q,P,\tau)$:
\begin{equation}
\frac{\partial S}{\partial q^a}=p_a\,,\qquad \frac{\partial
S}{\partial P^a}= Q_a\,,\qquad \frac{\partial
S}{\partial\tau}=-\mathcal{H}\,.
\end{equation}
The $P^a$ are constants of motion. If we therefore focus on the
appearance of the $q$'s we deduce from the Hamiltonian constraint
(${\cal H} = 2 c^2$)
\begin{align}
& S(q,\tau)={\cal W}(q) -2c^2\tau\,,\label{HJ1}
\end{align}
and the velocities follow a gradient flow set by $\cal W$:
\begin{equation}
p_a=\frac{\partial S}{\partial q^a}\qquad\Rightarrow\qquad
\dot{q}^a=G^{ab}\frac{\partial {\cal W}(q)}{\partial
q^b}\,.\label{HJ2}
\end{equation}
Combining (\ref{HJ1}) and (\ref{HJ2}), we have
\begin{equation}
\tfrac12 G^{ab}\partial_a {\cal W}^i\partial_b {\cal W}
+\calv(q)=2c^2\,. \label{HJ3}
\end{equation}
The function ${\cal W}$ is often called Hamilton's characteristic
function.  Note that the Hamilton--Jacobi equations always coincide exactly
with a rewriting of the action as a sum and difference of squares,
up to a total derivative:
\begin{equation}\label{sumofsquares}
S=\int \frac12  G_{ab}(\dot{q}^a - G^{ac}\frac{\partial {\cal W}}{\partial q^c}
)(\dot{q}^b - G^{bd}\frac{\partial {\cal W}}{\partial q^d} )\,.
\end{equation}

It is a basic theorem of analytical mechanics that, locally, one can
always define the first-order Hamilton--Jacobi equations. Typical
global phenomena are the existence of branch cuts such that $S$ can
become multi-valued. Barring these subtleties one can safely claim
that there always exists a function $S$, for any solution.

\section{Fake supersymmetry and Hamilton--Jacobi}\label{sect:4}

In this section we explain the link between Hamilton--Jacobi theory
and fake supersymmetry for domain walls and extremal black holes
(${\cal H}= 2 c^2 = 0$). The condition for fake supersymmetry is
that Hamilton's characteristic function factorizes as
$\mathcal{W}=\e^{aU}W(\phi)$.

We use this observation to make the statement that for
\emph{regular} extremal black holes, there is always a fake
superpotential. We prove this for large black holes, and conjecture
and motivate this for small black holes.

\subsection{Fake supersymmetry from Hamilton--Jacobi}
When the energy is zero (extremal black holes and domain
walls), Hamilton's characteristic function obeys (\ref{HJ3})
\begin{equation}
\frac{1}{4}(\partial_U \mathcal{W})^2 + \epsilon G^{ij}\partial_i
\mathcal{W}\partial_j \mathcal{W} = -\e^{2aU} V_{DW/BH}(\phi)\,.
\end{equation}
This relation suggests that there could be a simple factorised form
for $\mathcal{W}$
\begin{equation}\label{factorised}
\mathcal{W}=\e^{aU}W(\phi)\,.
\end{equation}
Exactly this assumption reproduces the (fake) supersymmetry flow equations
(\ref{floweqs}) and the defining relation for $W$ (\ref{Wequation})
from the Hamilton--Jacobi equations (\ref{HJ2}, \ref{HJ3}). The rewriting of the
action as a sum of squares (\ref{sumofsquares}) using Hamilton's
principal function then becomes the usual rewriting as a sum of
squares using the (fake) superpotential.

\subsubsection*{A new conserved quantity}
Let us make the condition for having a fake superpotential more
precise. Consider the effective action (\ref{effectiveaction})
with $c^2=0$ (Hamiltonian is zero on-shell). The Hamilton--Jacobi equations are
\begin{equation}\label{HJ}
4\dot{U}=\partial_U \mathcal{W}\,,\qquad \dot{\phi}^i= \epsilon
G^{ij}\partial_j \mathcal{W}\,.
\end{equation}
Now observe that
\begin{equation}
\mathcal{Q} \equiv \frac{4}{a}\dot{U} - \mathcal{W}\label{eq:Q_Is_Zero}
\end{equation}
is a constant of motion. This follows from:
\begin{align}
\frac{\d}{\d \tau}\mathcal{Q} &= \frac{4}{a}\ddot{U} - (\partial_i
\mathcal{W})\dot{\phi}^i - (\partial_U \mathcal{W})\dot{U}\nonumber\\
& = -\e^{2aU}V_{BH/DW}(\phi) - \epsilon
G_{ij}\dot{\phi}^i\dot{\phi}^j - 4\dot{U}^2
\end{align}
which is zero by virtue of the energy constraint ($\mathcal{H}=0$).
In the second line we used the Hamilton--Jacobi equations (\ref{HJ}) and the
$U$-equation of motion ($4\ddot{U}=-a\e^{2aU}V_{BH/DW}(\phi)$). It
follows that for non-extremal black hole solutions ($\mathcal{H}\neq
0$) this quantity is not conserved.

Now we consider flows for which
$\mathcal{Q}=0$. Then we can derive that
\begin{equation}
\partial_U \mathcal{W} = a\mathcal{W}\,,
\end{equation}
by the Hamilton--Jacobi equations (\ref{HJ}). If we integrate this
equation we find
\begin{equation}
\mathcal{W}=\e^{aU} W(\phi)\,,
\end{equation}
 We therefore find the elegant result
\begin{equation*}
\text{Fake SUSY}\qquad \Leftrightarrow \qquad \mathcal{Q}=0\,.
\end{equation*}
For non-extremal black holes ($c^2\neq 0$) it follows
straightforwardly that the factorisation property (\ref{factorised})
cannot hold, which was one of the central observations of
\cite{Perz:2008kh}. This is not that surprising since non-extremal
black holes can never be supersymmetric in any possible supergravity
theory.

There is a subtlety to the above statements. The principal
function $\mathcal{W}$ is by definition only determined up to a
constant. Therefore in principle we could set $\calq =0$ solution
by solution, by adding a moduli-dependent constant to $\calw$.
However, there is only fake-supersymmetry if the parameters can be chosen
so that the constant part of $\calq$ is solution-independent,
and it can be chosen to be \emph{identically} zero.
We clarify this point in the appendix, with some explicit
examples for which there is no solution-independent way of getting
$\calq=0$ and there is no fake supersymmetry.

\subsection{Fake supersymmetry and regularity}

Regular extremal black holes have an $AdS_2\times
S^2$ horizon. At this horizon one can show that $\mathcal{Q}=0$. For
that we use that all scalars are fixed at their attractor values
\begin{equation}
\phi (\tau=-\infty) =\phi_H\,,\qquad
\partial_{i}\mathcal{W}(\phi_H)=0\,.
\end{equation}
From the energy condition we then find that
\begin{equation}
\mathcal{W}(U,\phi_H) = 2\e^{U}\sqrt{-V_{BH}(\phi_H)}\,,
\end{equation}
such that $\mathcal{Q}=0$. Since $\mathcal{Q}$ is conserved it is
zero throughout the flow and the factorisation occurs everywhere.
\emph{Hence we have established that regular extremal flows must be
fake supersymmetric.} This is the usual assumption for the flow
equations of regular extremal black holes and here we provided a
proof that this is a necessary condition. Note that when $\calq=0$, the
asymptotic value of $\calw$ then gives the ADM mass through \eqref{eq:Q_Is_Zero}:
\be
\calw(\tau =0) = 4 M_{\rm ADM}\,.
\ee

A second class of extremal black holes that are still of interest
are the so-called extremal \emph{small} black holes. These are zero
entropy black holes, meaning they have vanishing horizon size.\footnote{In string theory,
these black hole develop a horizon through $\a'$ corrections.}
Equivalently, the black hole singularity is light-like. Since this
can be obtained by taking a limit of a regular solution we might
postulate similarly that also small extremal black holes are
necessarily fake supersymmetric. This limit is not a full
proof, and instead we make this into a conjecture for small extremal
black holes.

In summary, we make the following statement: Extremal flows
with $\calq= 0$ comprise all black holes with an $AdS_2 \times S^2$ horizon
(large black holes) and black holes for which the horizon
coincides with the singularity (small black holes). When $\calq\neq 0$,
the solution is unphysical and has one or more naked singularities (we give
such examples in the appendix). In the next section we will discuss some interesting
new features of small black holes in detail.

\section{Small black hole horizons}\label{sect:5}
For an extremal black hole with a macroscopic horizon area, the
attractor mechanism applies: the scalars at the horizon flow to
constants that are only functions of the charges. We prove that for
small black holes, a similar story holds. When fake supersymmetry is
valid, the scalars follow a geodesic flow near the horizon, even
though there is a scalar potential.

\subsection{Definition of a small black hole horizon}

The near-horizon geometry of a small black hole is conformal to $AdS_2 \times S^2$:
\be
ds^2_{\rm NH} = \rho^\a \left(R_{AdS}^2(-\rho^2 dt^2 +
\frac {d\rho^2}{\rho^2}) + R^2_{S^2}
\,d\Omega^2\right).\label{eq:AdS_Scaling}
\ee
The constant radii of $AdS_2$ and $S^2$ do not have to be equal. The conformal factor $\rho^\a$
is a simple consequence of
expanding the general conformal factor and keeping the leading term.
One can show that the constant $\a$ must be positive. To illustrate this,
consider the D0-D4 STU black hole: \be ds^2 = - e^{2U} dt^2 +
e^{-2U} (dr^2 + r^2 d \Omega^2)\,, \ee with warp factor \be e^{-2U}
= \sqrt{H_0 H^1 H^2 H^3}\,. \ee The four harmonic functions $H$ are
of the form: \be H^I = 1 + \frac {Q^I} r\,. \ee When one of the
charges is zero, the near-horizon geometry ($\rho \to 0$) is exactly
of the form \eqref{eq:AdS_Scaling}, with $\rho = \sqrt{r}$ and
$\a = 1$.

The small black hole near-horizon geometry can be interpreted as a
`scaling solution', similar to DW and cosmological scaling solutions
(see e.g.~\cite{Tolley:2007nq, Chemissany:2007fg}). The defining
property of a scaling solution is the existence of a conformal
Killing vector. Such a vector defines a local transformation that
preserves the metric up to a constant rescaling. For
the above metric \eqref{eq:AdS_Scaling} one can easily verify that the
transformation
\begin{equation}
\rho\rightarrow \e^{\lambda} \rho\,,\qquad t\rightarrow
\e^{-\lambda} t\,,
\end{equation}
leaves the metric invariant up to a \emph{constant} factor
\begin{equation}
g_{\mu\nu}\rightarrow \e^{\alpha\lambda} g_{\mu\nu}\,.\label{eq:ScalingMetric}
\end{equation}
It is useful to contrast this with large black holes. Large black
holes ($\alpha=0$) interpolate from Minkowski space at infinity to
$AdS_2\times S^2$ at the horizon. The general solution can therefore
be understood as a flow between two fixed points. The fixed points
themselves are characterized by a sudden increase in bosonic (and
sometimes fermionic) symmetries. For example, $AdS_2\times S^2$
realises the conformal group. Small black holes are similar in that
respect since, at the horizon, the solution is a scaling solution
and it is characterised by the same increase in bosonic symmetries,
the only difference is that the symmetries rescale the metric up to
a constant. For a recent discussion on the connection between
scaling solutions and black brane horizons we refer to
\cite{Chemissany:2011gr} and for its
applications to holography, see \cite{Kanitscheider:2008kd}.\footnote{From the extensive literature on
scaling solutions it is clear that scaling solutions occur indeed as
fixed point solutions to specific autonomous systems.}

\subsection{From scaling to Killing and geodesic flows}

For scaling solutions, the on-shell action scales with
on overall factor as well. Hence, an equivalent definition of
scaling solutions is that each term in the effective action scales
in the same way. What does this imply for the scalar fields? We
formalize this, following the strategy of \cite{Chemissany:2007fg}.

Consider the continuous transformation with parameter $\lambda$ such
that \be g_{\mu\nu} \to e^\lambda g_{\mu\nu}\,, \ee and the action
scales in the same way \be S \to e^\lambda S\,. \ee The small black
hole near-horizon geometries sit in this class. We recall the
argument of \cite{Tolley:2007nq} that the velocity field of the
scalars defines a Killing flow. The scaling of the action implies
that the velocity squared on the scalar tangent space is independent
of the scaling parameter $\lambda$: \be \frac{d}{d\lambda}
\left[G_{ij}\dot \phi^j\dot \phi^j\right] = 0\,. \ee If the
first-order derivative of the scalars with respect to $\lambda$
defines a vector field: (there is no explicit $\lambda$ dependence)
\be \xi^i\equiv \frac{d \phi^i}{d\lambda}(\phi)\,, \ee then we get
that the vector field $\xi$ is Killing since \be
(\call_\xi G_{ij}) \dot \phi^i \dot \phi^j=\frac{d}{d\lambda} \left[G_{ij}\dot \phi^i\dot \phi^j\right] =
0\,. \ee Hence we find that
for scaling solutions,  $\xi^i = \dot \phi^i$ is a Killing vector
field: \be \nabla_{(i} \xi_{j)} = 0\,. \ee The role of fake
supersymmetry now becomes clear. When there is a fake superpotential
\be \xi^i = G^{ij}
\partial_j W\,, \ee we find that also \be \nabla_{[i} \xi_{j]} =
\partial_{[i} \xi_{j]}=0\,, \ee and hence $\xi^i\nabla_i \xi_j=0$: the
scalars satisfy a geodesic equation.

This interesting connection between scaling solutions and geodesics
on moduli space was first made in \cite{Karthauser:2006ix}. If we
apply our conjecture that regular small black holes are always fake
supersymmetric, then we find that \emph{the scalar fields of regular
small black holes follow a geodesic flow near the horizon.}

Even though the action \eqref{BHeffectiveaction} has a scalar
potential, near the horizon of a small black hole the flow of the
scalars naively seems free since they describe a geodesic. This
``paradox'' gets resolved when one realises that the geodesic curve
is not an arbitrary geodesic. It is a very specific geodesic that is such
that $V_{BH}(\phi)$ scales in exactly the same way as the kinetic
term. Since a generic potential added to some sigma model cannot
have such a property, this simply means that scaling solutions
require special potentials, consistent with the fact that small
black holes are solutions to specific sets of charges.

\section{Discussion}\label{sect:6}

In this paper we analysed the difference between the first-order
Hamilton--Jacobi equations (which always exist), and the standard
fake supersymmetry equations. As we emphasized, not all extremal
black hole flows and domain wall flows are fake supersymmetric. To
our knowledge, the first example where this was shown appeared in
\cite{Sonner:2007cp}. This example describes an axion-dilaton domain
wall. We have given simpler domain wall and black hole examples in
the appendix, which are not fake supersymmetric.\footnote{Note that
domain wall examples of sections \ref{ssec:DW1}, \ref{ssec:DW2} are
the same, up to signs, as the black hole examples of sections
\ref{ssec:BH1}, \ref{ssec:BH2}. This was chosen on purpose to show
the similarity between domain wall flows and extremal black hole
flows.} A common feature of these examples is the presence of cyclic
scalars for which the momentum is a non-zero constant. It is
straightforward to demonstrate that this is the reason why there is
no fake superpotential. The argument proceeds as follows. Denote the
cyclic scalar as $\phi_c$, its conjugate constant momentum as $p_c$
and the other scalars as $\phi^i$. From the Hamilton--Jacobi
equation
\begin{equation}
p_c=\partial_{\phi_c}\calw(\phi_c, \phi^i, U) \,,
\end{equation}
we find that
\begin{equation}
\mathcal{W}(\phi_c, \phi^i, U) =  \phi_c p_c +
\tilde{\mathcal{W}}(U,\phi)\,,
\end{equation}
and is hence never of the factorised form $\mathcal{W}\neq
\e^{aU}W(\phi_c,\phi^i)$.

This does not necessarily imply that all solutions that fail to be
fake supersymmetry are necessarily having cyclic scalars. That is why we
formalised the condition for having fake supersymmetry in an elegant
condition: a certain conserved charge $\mathcal{Q}$ has to vanish.
This conserved charge is given by
\begin{equation}
\mathcal{Q} =\frac{4}{a}\dot{U} - \mathcal{W}\,,
\end{equation}
with $\mathcal{W}$ Hamilton's characteristic function. Since we
wrote down a very general Hamiltonian system (\ref{metric},
\ref{potential}, \ref{effectiveaction}) it is non-trivial that one
can demonstrate the existence of a universal conserved quantity,
different from the energy. The fact that we were able to do this is
because our Hamiltonian system is not completely arbitrary, the
warp factor $U$ appears in a specific way in the metric on
configuration space (\ref{metric}) and it factors out in the potential
(\ref{potential}).

With this precise formulation of fake supersymmetry (versus
Hamilton--Jacobi) we were able to derive a physical meaning of fake
supersymmetry for extremal black holes: it is a necessary condition
for having physically acceptable solutions. For example, the models
with cyclic scalars with non-zero momentum have the same kind of
unphysical behavior of the black hole warp factor as for
over-extremal solutions\footnote{A prototypical example of this
phenomenon would be extremal black holes in $\mathcal{N}=2$
supergravity with non-constant hypermultiplet scalars.}.  We have
presented a rigorous proof of this for extremal black holes that
have a finite horizon (large black holes), and we have argued, by
taking a limit to vanishing horizon size, that the same applies to
small black holes. However it would be more satisfying to have a
rigorous proof for small black holes as well. Perhaps a useful
playground to test this conjecture for small black holes is to
investigate solutions in supergravity theories with symmetric scalar
manifolds. Then there is full understanding of the space of
solutions \cite{Breitenlohner:1987dg}, simple integration algorithms
have been developed \cite{Chemissany:2010zp} and the link with the
first-order formalism has been investigated in detail, see for
example \cite{Chemissany:2010zp, Bossard:2009we}.

We furthermore pointed out that the near-horizon geometry of small
black holes is of a universal ``scaling form'', which means it is
characterised by an increase in symmetries that do not preserve the
metric but rescale it with a constant factor. Using this insight, we
have borrowed techniques from scaling solutions in cosmology
\cite{Tolley:2007nq, Karthauser:2006ix, Chemissany:2007fg} to
uncover a general pattern in the behavior of the scalar fields when
they flow towards the horizon of a small black hole. Instead of
reaching specific constants, as they do for large black holes, the
scalars start to follow a specific geodesic curve on moduli space.
This is counter-intuitive since the scalars are subject to a non-zero
potential. But just as for scaling attractors in cosmology one has to
restrict to specific geodesic curves that are such that the
potential scales in the same way as the scalar kinetic term. This is
the extension of the well-known attractor mechanism for large black holes:
the behavior of the scalars near the horizon is dictated by the
increase in symmetry (scaling) and determined by the charges only
(the geodesic is of a specific kind). It should be possible to generalise these  results
to general black $p$-branes in $D$ dimensions, along the lines
of \cite{Martin:2012bi}.

If we come back to the main motivation for studying fake supersymmetry, which
is finding first-order equations of motion to facilitate the search
for solutions, then there is clearly no reason to worry about the
existence of a fake superpotential, since the first-order
Hamilton--Jacobi equations always exist locally. However,
generically the problem of finding Hamilton's principal function
that governs the first-order HJ flow equations is as difficult as
integrating the second-order equations once. So from that point of view
there is not much technical gain. What counts, in our opinion, is the
physical meaning of having fake supersymmetry. For domain walls, it
guarantees stability of the solution \cite{Freedman:2003ax},
and as we showed in this paper, for extremal black holes it is a
necessary condition for having physically sound solutions.

It would be interesting to apply the above insights to the
Hamiltonian system defined by the so-called baryonic branch of the
Klebanov--Strassler background \cite{Butti:2004pk}. For this system
there seems an unsettled issue about the existence of first-order
gradient flow equations \cite{Halmagyi:2011yd, Giecold:2011kf}.

\section*{Acknowledgments}
TVR and BV are supported respectively by the ERC Starting
Independent Researcher Grant 259133-ObservableString and
240210--String--QCD--BH.

\appendix

\section{Examples} \label{sec:Examples}

In this section, we consider some explicit examples to clarify the
notions of the previous sections. We focus on domain wall solutions
and extremal black holes, for two simple types of effective
Lagrangians: the metric scalar $U$ with its potential and a free
scalar, and the metric scalar coupled to an axion-dilaton system,
with a potential that depends on $U$ and the dilaton.

\subsection{Domain wall example 1: \texorpdfstring{$\Lambda$}{} plus free scalar}\label{ssec:DW1}
Consider gravity coupled to a negative cosmological constant
$\Lambda$ and a free scalar $\phi$. The effective action is simply
\begin{equation}
S=\int \d z\Bigl( 4\dot{U}^2 -\dot{\phi}^2 - \e^{2\sqrt3 U}\Lambda
\Bigr)\,.
\end{equation}
The solution for $\phi$ is straightforward
\begin{equation}\label{integration1}
\phi(z)=p\,z + \phi(0)\,,
\end{equation}
with $p$ a constant, describing the scalar's momentum. Without loss
of generality we consider it to be non-negative. From the energy
constraint we can integrate the equation for $U$ once
\begin{equation}\label{integration2}
\dot{U}=\pm \frac{1}{2}\sqrt{p^2 -\e^{2\sqrt3\, U}\Lambda}\,.
\end{equation}
By redefining $z$ we can always fix the sign above. We take the
plus sign. The explicit solution is then given by
\begin{equation}
\e^{-\sqrt3 U} =
\frac{\sqrt{-\Lambda}}{|p|}\sinh(-\tfrac{\sqrt3}{2}|p|\, z)\,.
\end{equation}
When $p=0$ we find
\begin{equation}
\e^{-\sqrt3 U} = -\frac{\sqrt{-3 \Lambda}}{2}\, z\,,
\end{equation}
and after the coordinate transformation $z =
-\frac{\sqrt{-3\Lambda}}{2}\, \rho^3$, we find the $AdS_4$ metric in
standard Poincar\'e coordinates.

To find $\mathcal{W}(U,\phi)$ for the general solution we use the
above two integrations (\ref{integration1},\ref{integration2})
\begin{equation}
\partial_{\phi}\mathcal{W} = -p\,,\qquad \partial_U \mathcal{W} = 2
\sqrt{p^2 -\e^{2\sqrt3\, U}\Lambda}\,,
\end{equation}
such that
\begin{equation}
\mathcal{W}(U,\phi) = -p\phi  -2p\,U+ \frac{2}{\sqrt3}\sqrt{p^2
-\Lambda\e^{2\sqrt3 U}}   + \frac{2p}{\sqrt3}\ln\bigl(\sqrt{
p^2-\Lambda\e^{2\sqrt3 U}} -p \bigr)\,.
\end{equation}
When $p=0$ we indeed find the factorised form
\begin{equation}
\mathcal{W}(U,\phi) =  \frac{2}{\sqrt3}\sqrt{-\Lambda}\e^{\sqrt3
U}\,.
\end{equation}
When $p\neq0$, $\calw$ does not factorize and there is no fake supersymmetry
in the standard sense.

The constant of motion $\mathcal{Q}$ is
\begin{equation}
\mathcal{Q} = p\,\phi(0) - \frac{p}{\sqrt3}\ln(-\Lambda)\,.
\end{equation}
It is clearly moduli dependent for $p\neq 0$ (appearance of $\phi(0)$). Following
the arguments of section \ref{sect:4}, we cannot set $\calq=0$ in a solution-independent
way  and this explains why $\calw$ does not factorize.

\subsection{Domain wall example 2: the Sonner--Townsend model}\label{ssec:DW2}
The first known example of a domain wall (and cosmological)
solution, that can not be derived from a fake superpotential in the
usual sense, was found by Sonner and Townsend in
\cite{Sonner:2007cp}. The domain wall solution is a scaling solution
and therefore it represents a fixed point of a more general domain
wall flow, which is not known analytically. We briefly
review this solution and simplify its presentation.

The essential ingredient is again a free scalar, but this time it is
an axionic field $\sigma$ that couples to the dilaton $\phi$, as
follows
\begin{equation}
\mathcal{L} = 4\dot{U}^2 -\dot{\phi}^2 - \e^{\mu\phi}\dot{\sigma}^2
- \Lambda\e^{2\sqrt3 U + \lambda\phi}\,,
\end{equation}
where the constants $\mu,\lambda$ define the couplings of the model.
The target space spanned by $\phi$ and $\sigma$ is the coset $SL(2,
\Real)/SO(2)$. The $SL(2,\Real)$ symmetry of the kinetic term is
broken by the dilaton potential. Since the domain wall solution is a
scaling solution, the scalars follow a Killing flow, but the lack
of a fake superpotential implies that the Killing flow is not a geodesic
flow, as explained in the section \ref{sect:5}.

The scaling Ansatz is
\begin{equation}
U = a\ln z + U_0\,,\qquad \phi = b\ln z +\phi_0\,,
\end{equation}
where $a, b, U_0, \phi_0$ are constants. The axion equation of
motion implies
\begin{equation}
\dot{\sigma}=\frac{d}{ z^{\mu b}}\,,
\end{equation}
with $d$ a constant. If we furthermore demand the scaling condition,
which means that the $\sigma$-kinetic term scales similar to the
other terms in the actions, we can fix $b$
\begin{equation}
b =\frac{2}{\mu}\,.
\end{equation}
The $U$ and $\phi$ equations of motion lead to three algebraic
relations amongst the four remaining integration constants $a, d,
U_0, \phi_0$:
\begin{align}
& a= -\frac{1}{\sqrt3}(1+\frac{\lambda}{\mu})\,,\\
& d^2 =4\e^{-\mu\phi_0}\Bigl( -\frac{1}{\mu^2}
+\frac{\lambda}{3\mu}(1 +
\frac{\lambda}{\mu})\Bigr)\,,\\
&4(1+\frac{\lambda}{\mu}) + 3\Lambda\e^{2\sqrt3U_0  +
\lambda\phi_0}=0\,.
\end{align}
The Hamiltonian constraint is automatically fulfilled with these
relations. These relations imply certain sign restrictions on the
possible choices for $\mu,\lambda, \Lambda$, which we do not
discuss.

As in the previous example we can  show that a fake
superpotential ($\calw$ factorizes as $\calw = e^{\sqrt 3 U} W(\phi,\sigma)$) requires one of the integration constants to be zero
($d=0$) and the system collapses to
the single-dilatonic domain wall flow. The expression for
$\mathcal{W}$ for the general solution with $d \neq 0$ is quite
involved and not that insightful.

\subsection{``Black hole'' example 1: Maxwell plus free scalar}\label{ssec:BH1}
Consider Einstein--Maxwell theory with a free scalar added to it.
Electric solutions have the following expression for the field
strength $F = \d (\chi(\tau)\d t)$ where $\chi$ is the
electric potential. Its equation of motion is $ \e^{-2U}\dot{\chi} = q$,
where $q$ is the electric charge. This means
that the black hole effective potential $V_{BH}$ is simply a
constant $V_{BH}=-q^2$. The effective action is
completely analogous to that of section \ref{ssec:DW1}:
\begin{equation}
S=\int \d z\Bigl( 4\dot{U}^2 +\dot{\phi}^2 + q^2\e^{2U}\Bigr)\,.
\end{equation}
The solution for $\phi$ is
\begin{equation}
\phi(\tau) =p \tau +\phi(0)\,,
\end{equation}
where $p$ is a constant which we take to be non-negative without
loss of generality. From the energy condition we have that,
\begin{equation}
2\dot{U} =\pm\sqrt{q^2\e^{2U} - p^2}\,.
\end{equation}
The choice of plus and minus sign is a gauge fixing and will
determine how the radial variable $\tau$ is related to the usual
radius $r$. The solution (for the plus sign) is
\begin{equation}
\e^{-U} = \frac{|q|}{p} \sin(\tfrac{1}{2}\,p\,\tau)\,.
\end{equation}

Let us  verify that $p=0$ gives the standard electric extremal RN
black hole. The solution for $U$ becomes
\begin{equation}
\e^{-U}=-\frac{|q|}{2}\tau +1\,,
\end{equation}
where we fixed an integration constant to be 1 for simplification.
The coordinate transformation $r =\pm (\frac{1}{\tau}
-\frac{|q|}{2})$, gives the following metric
\begin{equation}
\d s^2 = - (1 + \frac{|q|}{2r})^2\d t^2 + (1 +
\frac{|q|}{2r})^{-2}\d r^2 + r^2 \d\Omega^2\,,
\end{equation}
which we indeed recognize as the standard text book expression for
the extremal RN black hole.

From the first-order equations we can find $\mathcal{W}$ to be
\begin{equation}
\mathcal{W}(\phi, U) = p\,\phi  + 2\sqrt{q^2\e^{2U} - p^2} -
2p\arctan(\frac{1}{p}\sqrt{q^2\e^{2U}-p^2})\,.
\end{equation}
In the limit $p\rightarrow 0$ this consistently reduces to
\begin{equation}
\mathcal{W}(\phi, U) = 2q \e^{U}\,.
\end{equation}
For $\mathcal{Q}$ we find
\begin{equation}
\mathcal{Q}=p\,\phi(0) -2pk\pi\,,\qquad k \in \mathbb{Z}.
\end{equation}
This is only manifestly zero when $p=0$. When $p\neq0$ it depends on the modulus at infinity $\phi(0)$
and there is no fake supersymmetry.

It is not difficult to verify that the solution has a naked
singularity. Moving from $\tau=0$ at spatial infinity to finite
$\tau$ we hit a singularity of the metric at $\tau=\frac{2
\pi}{|p|}$. Note that this metric singularity is not lightlike (it is
not a horizon). One can easily verify that the curvature invariant
$R_{abcd}R^{abcd}$ blows up near  the metric
singularity.%

\subsection{``Black hole'' example 2: Dilatonic black hole plus axion}\label{ssec:BH2}
For black hole flows we can have the exact analogy with the
Sonner--Townsend domain wall scaling solution of section \ref{ssec:DW2}. This can be
engineered by extending the usual dilaton black hole solution with
an axion $\sigma$. The action is

\begin{equation}
\mathcal{L} = 4\dot{U}^2 +\dot{\phi}^2 + \e^{\mu\phi}\dot{\sigma}^2
- q^2\e^{2 U + \lambda\phi}\,.
\end{equation}
The derivation follows exactly the same rules as with the domain
wall. The scaling Ansatz is again
\begin{equation}
U = a\ln \tau + U_0\,,\qquad \phi = b\ln \tau +\phi_0\,,\qquad
\dot{\sigma}=\frac d {\tau^{2}}\,.
\end{equation}
Where the equations of motion imply the following relations between the integration constants:
\begin{align}
& a= -(1+\frac{\lambda}{\mu})\,,\\
& b =\frac{2}{\mu}\,,\\
& d^2 =-4\e^{-\mu\phi_0}\Bigl( \frac{1}{\mu^2} +
\frac{\lambda}{\mu}(1 + \frac{\lambda}{\mu})\Bigr)\,,\\
&4(1+\frac{\lambda}{\mu}) + q^2\e^{2U_0  + \lambda\phi_0}=0\,.
\end{align}
This is, as far as we know, a new solution, but by itself does not
represent a black hole solution since it is not asymptotically flat.
Being a scaling solution one might expect that it describes the near horizon
region of a small black hole. However, our general theorem about the
absence of fake supersymmetry  implies that this solution cannot be
interpreted that way. If one extends the solution to the general
solution that interpolates to this scaling solution, one will
encounter naked singularities in the bulk.

\providecommand{\href}[2]{#2}\begingroup\raggedright\endgroup

\end{document}